\documentclass[12pt]{iopart}
\usepackage{graphicx}

\begin{document}

\title[Effects of income redistribution on the evolution of cooperation]{Effects of income redistribution on the evolution of cooperation in spatial public goods games}

\author{Zhenhua Pei$^1$, Baokui Wang$^2$\footnote{Author to whom any correspondence should be addressed.} and Jinming Du$^3$}
\address{$^1$ Department of Prosthodontics, Capital Medical University School of Stomatology, Beijing, 100050, China}
\address{$^2$ Unit No. 92060 of PLA, Dalian, 116041, China}
\address{$^3$ College of Information Science and Engineering, Northeastern University, Shenyang, 110819, China}
%\address{$^5$ Author to whom any correspondence should be addressed.}
\ead{\mailto{peizhenhua1983@163.com}, \mailto{baokuiwang@outlook.com} and \mailto{jmdu1987@163.com}}
\vspace{10pt}
\begin{indented}
\item[]October 2016
\end{indented}

\begin{abstract}
Income redistribution is the transfer of income from some individuals to others directly or indirectly by means of social mechanisms, such as taxation, public services and so on.
Employing a spatial public goods game, we study the influence of income redistribution on the evolution of cooperation.
Two kinds of evolutionary models are constructed, which describe local and global redistribution of income respectively.
In the local model, players have to pay part of their income after each PGG and the accumulated income is redistributed to the members.
While in the global model, all the players pay part of their income after engaging in all the local PGGs, which are centered on himself and his nearest neighbours, and the accumulated income is redistributed to the whole population.
We show that the cooperation prospers significantly with increasing income expenditure proportion in the local redistribution of income, while in the global model the situation is opposite.
Furthermore, the cooperation drops dramatically from the maximum curvature point of income expenditure proportion.
In particular, the intermediate critical points are closely related to the renormalized enhancement factors.
\end{abstract}

\pacs{02.50.Le, 87.23.Kg, 87.23.Ge}
\noindent{\it Keywords\/}: evolutionary game theory, public goods game, human cooperation, income redistribution

\submitto{\NJP}
\maketitle

\section{Introduction}
%%public goods
The stability of most complex systems, ranging from human societies to animal kingdoms, relies on public goods that individuals collected.
The most precise technical definition of a public good by {\it Samuelson} is a good that can be consumed by an additional consumer without additional cost\cite{Samuelson1954}.
However, in the consciousness of most people, the term conjures the image of a good available for all citizens to consume, such as national defense, environmental protection, health insurance and highways\cite{Holcombe1997}.
Most governments devote considerable resources to the provision of private goods, and such universal provision schemes can
redistribute income from the rich to the poor\cite{Besley1991}.
%%Redistribution of income
Redistribution of income may provide a nonexcludable benefit to those who give, and many such schemes are universal in the sense that everyone is eligible and the provision is free\cite{Blackorby1988}.
One of the classic forms of income redistribution is the tax system, in which people are taxed at fixed rates.
People who make more money pay higher taxes, thereby forfeiting more of their income to the government.
Tax funds are used to benefit the society as a whole by providing a variety of public and social services by the government, and the direct transfer of income may occur in the case of welfare payments and other forms of cash assistance made to low-income members of society.

%%cooperation
Although natural selection favours the fittest and most successful individuals, which in turn implies an innate selfishness that greatly challenges the concept of cooperation, economic impact on the evolution of cooperation in human societies may just be the missing ingredient for cooperative behaviour to prevail\cite{Szolnoki2016}.
%%evolutionary game theory
As the theoretical foundation in this realm, evolutionary game theory has become one of the most frequently-used approaches to understand how cooperation prevails in social-economic systems\cite{Nowak2012,Szolnoki2013}.
Such kinds of systems in which successful strategies spread by imitation or genetic reproduction are routinely analyzed in evolutionary biology, sociology, anthropology and economics.
Recently, the application of methods from statistical physics to these systems has led to many important insights.
%%public goods game
It is known that public goods game (PGG) is one of the most famous paradigms in evolutionary game theory, which is played by groups and often used for discussing the conflict between individuals and common interests\cite{Rand2013}.
In a typical PGG, $n$ players are asked to decide whether or not to contribute to a common pool.
They are clear that the total amount of investments will be multiplied by an enhancement factor $r (r>1)$ and then be distributed equally to all the players irrespective of their contributions.
Apparently, players who do not contribute fare better than contributors of the group.
Consequently, the whole system may evolve towards the ``tragedy of the commons''\cite{Hardin1968}.

%%some useful previous work
The emergence of cooperation in sizable groups between unrelated individuals puzzles diverse fields of social sciences\cite{Wang2014}.
Many mechanisms have been proposed to elucidate the prosperity of cooperative behaviours under the exploitation of defectors in PGG, such as voluntary\cite{Hauert2002}, punishment\cite{Brandt2003,Helbing2010,Chen2014}, social diversity\cite{Santos2008}, heterogeneous wealth distribution\cite{Wang2010}, diverse contribution\cite{Gao2010}, Matthew effect\cite{Perc2011}, coevolution\cite{Wu2009,Zhang2011} and interdependent networks\cite{Wang2012,WangZ2014}.
%%characteristics of this work
However, it is known that human altruism goes far beyond that which has been observed in the animal world\cite{Fehr2003}.
In the social-economic system, it is generally agreed that giving is motivated by a variety of factors other than altruism\cite{Andreoni1988}.
Few of previous researches on physical models has considered how specific factors, such as redistribution of income, influence the evolution of cooperation in the complex social-economic system.

%%Our work
Inspired by the seminal works, we consider diverse levels of redistribution of income in the spatial PGG.
Concretely, the effects of local and global redistribution of income on the evolution of cooperation are studied respectively in this work.
Basically, we assume that each player engages in the local PGGs, which are centered on himself and nearest neighbours to obtain their local income.
Based on this, in the model of local redistribution of income, players have to pay part of their idealized income to the  focal group according to a given income expenditure proportion after each round.
%%That is, the game process of each PGG between a player and his nearest neighbours is considered as a transaction process in one group.Players have to pay part of their incomes to the public pool according to a fixed ratio if they are positive.
Distinguished from the contribution action during the PGG process, such compulsory payment is named as the second-order payment.
Subsequently, the accumulated income are redistributed to all the members of this group regardless of their strategies and the quantity of their second-order payments.
On the other hand, in the global redistribution of income, players pay part of their idealized income to the whole population.
Similarly, the accumulated income is then redistributed to all the players in the whole population.
However, the two types of redistribution models are essentially different in our work, which play different roles in different stages in spatial PGG.
In reality, the local redistribution of income seems like a special transaction tax in economic system, which is collected according to the definite quantity of the volume of trade.
And then, the revenue is redistributed to the group members uniformly, which amounts to the fiscal subsidy for a particular industry.
While in the global redistribution of income, the processes of second-order payment and income redistribution can be classified as the process of collecting and redistributing the gross income of personal income tax for the whole population.
Apparently, the processes of collection and redistribution have essential differences between two types of income redistribution models.
Remarkably, we find that the evolution of cooperation in spatial PGG is significantly promoted with the increasing income expenditure proportion of the whole population in local redistribution of income.
While in global redistribution of income, the evolutionary trends of cooperation is opposite.
Meanwhile, there exists a intermediate critical point of income expenditure proportion for each fixed renormalized enhancement factor, where cooperative behaviours drop dramatically.

\section{Spatial PGG with local redistribution of income}
%%square lattice
We employ one $L{\times}L$ square lattice with periodic boundary condition and von Neumman neighbourhood, where each player is surrounded by $M=4$ local nearest neighbours and the group size of each PGG is $G=M+1$.
There are no empty sites on the lattice, and each player engages in $G$ local PGGs which are centered on himself and the nearest neighbours.
%%why employ square lattice
It is noted that the square lattice is the simplest network that allows us to go beyond the well-mixed population assumption and take into account the interactions engaged in group interactions among players rather than random\cite{Szolnoki2016}.
Also, it allows us to investigate the effects of redistribution of income on the evolution of cooperation more detailed without the influence of heterogeneity.
%%PGG
Initially, players are designed either as a cooperator or a defector with equal probability.
Without loss of generality, cooperators contribute 1 to the common pool and defectors contribute nothing in each PGG.
The total contribution is subsequently multiplied by an enhancement factor $r$, and then equally distributed to the $G$ members in the same group irrespective of their strategies.
Consequently, the income of defectors engaging in one PGG is $\Pi_d=r\cdot{n_c}/G=\eta\cdot{n_c}$, and the corresponding income of cooperators $\Pi_c=\Pi_d-1$, where $n_c$ is the number of cooperators in the group and $\eta=r/G$ denotes the renormalized enhancement factor in each PGG.

After game interaction, each player has to pay part of their income to the group according to the given income expenditure proportion $p_{lr}$.
We emphasize that parameter $p_{lr}$ denotes the proportion of the income obtained in the single PGG of the focal group.
Subsequently, the accumulated income expenditure of players in this group is redistributed to the $G$ group members irrespective of their strategies and the amount of their second-order payments.
Thus, the actual income of each player in this group after local redistribution of income is
\begin{equation}
\label{PayoffAfterLRI}
\Pi_{li}'=\Pi_{li}\cdot(1-p_{lr})+S_{lG}/G,
\end{equation}
where $\Pi_{li}$ denotes the income of a player after one PGG, $S_{lG}=p_{lr}{\cdot}\sum^G_{li=1}\Pi_{li}$ is the sum of the accumulated second-order payments of the group members.
We assume that, when a single public good is provided at positive levels by private individuals, its provision is free and unaffected by a redistribution of income\cite{Warr1983}.
In addition, we assume that the redistribution process is no cost and the player pays nothing to the group if his income $\Pi_{li}$ is not positive after the corresponding PGG.
After engaging in all the groups centered on himself and nearest neighbours, a player obtains the final payoff $\Pi_l=\sum^G_{li=1}\Pi_{li}'$.
Subsequently, each player is allowed to learn from one randomly selected nearest neighbour.
To be specific, player $x$ adopts the strategy of the random neighbour $y$ with a probability determined by the difference of their final payoffs
\begin{equation}
\label{Fermi}
W_{(x{\leftarrow}y)}=\frac{1}{1+\exp[(\Pi_{x}-\Pi_{y})/{\kappa}]},
\end{equation}
where $\kappa$ denotes the selection intensity in strategy update process.
Following a previous study\cite{Szabo1998}, we simply set $\kappa=0.5$ in this work and mainly focus on the effects of income redistribution on the evolution of cooperation in spatial PGG.
Also in this work, we adopt the synchronous Monte Carlo Simulation (MCS) procedure to update the strategies of players.
In addition, we employ the square lattice having $L=100$ linear size to avoid finite size effects.
Unless otherwise stated, all the simulation results shown in this section are required up to $10^4$ generations and then sampled by another $10^3$ generations.
The results of fractions of cooperators are averaged over 50 different realizations of initial conditions.

\begin{figure}
\begin{center}
\includegraphics[scale=0.4]{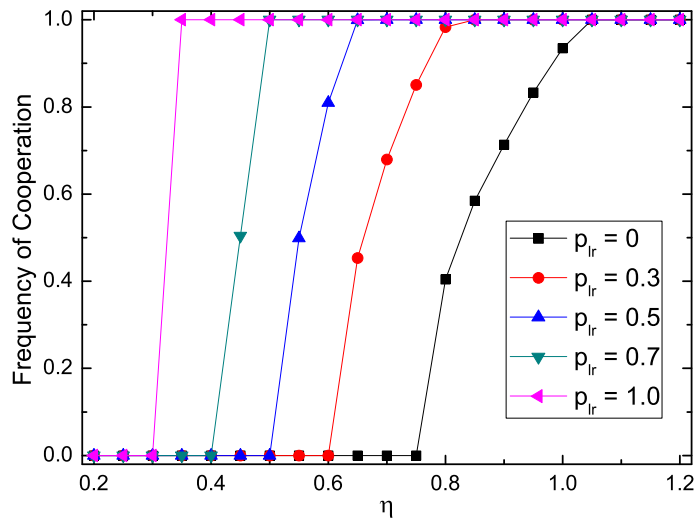}
\caption{The evolution of cooperation in spatial PGG with local redistribution of income as a function of $\eta$ with different values of $p_{lr}$.}
\label{LocalPlrs}
\end{center}
\end{figure}

In the following, we present the simulation results of the effect of local income redistribution on the evolution of cooperation in spatial PGG.
We first show the evolution of the cooperative behaviours of the whole population as a function of the renormalized enhancement factor $\eta$ with different values of $p_{lr}$ in \Fref{LocalPlrs}.
As shown in \Fref{LocalPlrs}, the evolution of cooperation on square lattice prospers dramatically with the increment of the income expenditure proportion $p_{lr}$, which helps the cooperative behaviours prevail under much worse conditions.
In particular, with the increment of $p_{lr}$ until $p_{lr}=1.0$, the survivability of cooperators on square lattice is enhanced steadily, and the whole population can easily achieve the full cooperation even at $\eta=0.35$.
It is well-known that, in the context of PGG, small values of enhancement factor favour defectors and large values benefit cooperators\cite{Hauert2003}.
In our work, compared with the frequencies of cooperation on $p_{lr}=0$, cooperators have a much better chance for survival when $p_{lr}>0$.
Qualitatively, after each PGG in a group, the income of each defector is $\Pi_{ld}$ and the corresponding income of each cooperator is $\Pi_{lc}$, where $\Pi_{ld}-\Pi_{lc}=1$.
Subsequently, after paying part of their income to the group according to the given probability $p_{lr}$, the rest income of each defector is $\Pi_{ld}'=\Pi_{ld}\cdot(1-p_{lr})$, and $\Pi_{lc}'=\Pi_{lc}\cdot(1-p_{lr})$ for the rest income of each cooperator.
Then, the accumulated income is redistributed to the group members irrespective of their strategies and contributions.
Accordingly, the final payoff of each defector in this group is $\Pi_{ld}''=\Pi_{ld}'+S_{lG}/G$, and $\Pi_{lc}''=\Pi_{lc}'+S_{lG}/G$ for the final payoff of each cooperator.
Consequently, the payoff difference between a cooperator and defector in one group is $\Pi_{ld}''-\Pi_{lc}''=\Pi_{ld}'-\Pi_{lc}'=(\Pi_{ld}-\Pi_{lc})\cdot(1-p_{lr})=1-p_{lr}$.
Thus, the payoff gap between cooperators and defectors is decreased with increasing $p_{lr}$.
That is, the mechanism of local redistribution of income narrows the wealth gap between cooperators and defectors, which makes the cooperation behaviours have more chance to prevail in the spatial PGG rather than without such mechanism.
Moreover, it is worth noting that each player engages in $G$ local PGGs which are centered on himself and nearest neighbours.
Local redistribution of income further balances the income difference of defectors and cooperators in one group.
Suffered as a consequence, the final payoff of each player on the square lattice is strongly dependent on the quality of the groups around\cite{Perc2013}.
Obviously, more cooperators make larger contributions in the group with a fixed number of participants.
Thereinto, a player surrounded by the groups with more cooperators has a competitive advantage over the neighbour players around.
Thus, the redistribution of income, which acts as a driving force for promoting cooperation on the square lattice, gives prominence to the role of groups on the evolution of cooperation and leads to persistent of full cooperation state over a wide range of parameter $p_{lr}$\cite{Bottcher2016,Su2016}.

\begin{figure}
\begin{center}
\includegraphics[scale=0.6]{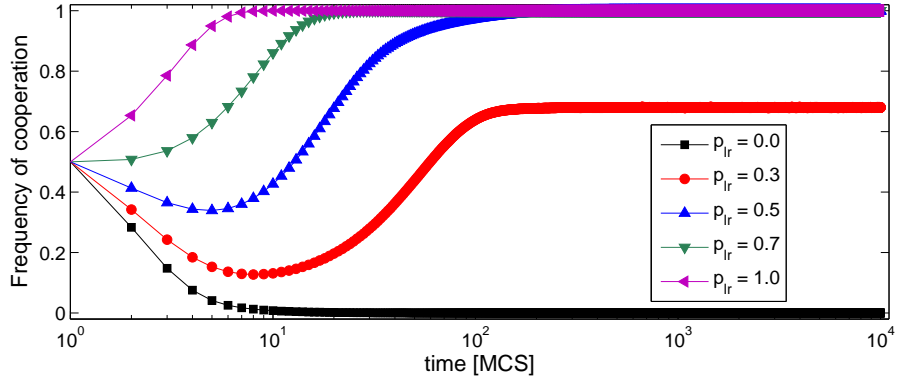}
\caption{
Time evolution as characterized by the global density of cooperators with different values of $p_{lr}$.
Initially, cooperators and defectors are randomly distributed on the square lattice with equal probability.
The results are shown for $\eta=0.7$.
The evolutionary processes are averaged over 1000 different initial conditions.
}
\label{LocalPlrTime}
\end{center}
\end{figure}

To investigate how the local redistribution of income affects the evolution of cooperation in spatial PGG intuitively, we present the time evolution of cooperation in \Fref{LocalPlrTime}.
Compared with the classical result, where $p_{lr}=0$, the time evolution of cooperation level prospers steadily with increasing $p_{lr}$.
It is known that cooperators can form compact clusters to prevent the invasion of defectors on square lattice\cite{Szolnoki2009}.
These compact clusters help cooperators acquire an evolutionary advantage and thrive gradually until reaching the steady states.
While in our model, there implies a propagation mechanism of evolutionary advantage for cooperation due to the local redistribution of income.
On one hand, the evolutionary advantage of defectors over cooperators decreases by reducing the income disparity between them with increasing $p_{lr}$.
Consequently, the motivation of learning for cooperators from defectors diminishes gradually.
On the other hand, players within the groups of more cooperators has evolutionary advantage over the others.
In particular, in the context of $p_{lr}\rightarrow1$, there is almost no payoff difference between cooperators and defectors within one group after the redistribution of income.
The only difference of performance exists among groups.
More cooperators make more benefits in one PGG, which help the group members obtain more after the redistribution of income.
A player in the group with less cooperators inclines to adopt the strategy of a random neighbour within the group including more cooperators spontaneously\cite{Pinheiro2012}.
Hence, the group including more cooperators enjoys the evolutionary advantage of cooperation, and makes the cooperative behaviours spread more rapidly on the square lattice with increasing $p_{lr}$.
From another point of view, the mechanism of local redistribution of income amplifies the differences between groups, which increases the heterogeneity on the square lattice and boosts the evolution of cooperation for the entire range of $p_{lr}$\cite{Santos2012}.
Thus, the redistribution of income directly affects the evolutionary dynamics of cooperation on the square lattice, which helps cooperators reproduce and survive much better, even in the harsh environment.

\section{Spatial PGG with global redistribution of income}
In this section, we describe the evolution of cooperation in spatial PGG with global redistribution of income in detail.
Similarly with the former evolutionary model with local redistribution of income, $L{\times}L$ square lattice with periodic boundary condition and von Neumman neighbourhood is employed, where no empty sites are allowed.
Players also engage in $G$ local PGGs which are centered on himself and nearest neighbours to gain their income.
Nevertheless, the process of income redistribution performs after completing the $G$ PGGs.
Therein, players are mandatory to pay the part of their income to the whole population according to a fixed probability $p_{gr}$, which is also named as the second-order payment as mentioned before.
We emphasize that $p_{gr}$ denotes the income expenditure proportion of all the income player obtained in the $G$ PGGs.
Subsequently, the accumulated income is redistributed to all the players irrespective of their strategies or the quantity of their second-order payments.
Thus, the final payoff of each player on the lattice is
\begin{equation}
\label{PayoffAfterGRI}
\Pi_{gi}'=\Pi_{gi}\cdot(1-p_{gr})+S_{gG}/(L{\times}L),
\end{equation}
where $\Pi_{gi}$ denotes the income of a player after $G$ PGGs, $S_{gG}=p_{gr}{\cdot}\sum^{L{\times}L}_{gi=1}\Pi_{gi}$ the accumulated second-order payments of all the players, $\Pi_{gi}'$ the final payoff of a player after the global redistribution of income.
Also, the redistributing process is no cost and the players pay nothing to whole population if $\Pi_{gi}$ is not positive after $G$ PGGs.
Similarly, we employ \Eref{Fermi} and set $\kappa=0.5$ in the update process.
We also adopt $L=100$ linear size to avoid finite size effects and the synchronous MCS procedure to update the strategies of players.
Except for the difference that, the simulation results shown below are required up to $3{\times}10^4$ generations and then sampled by another $10^3$ generations to make sure that the evolution of cooperation approaches an evolutionary stable state.
The results of fractions of cooperators are also averaged over 50 different realizations of initial conditions.

\begin{figure}
\begin{center}
\includegraphics[scale=0.4]{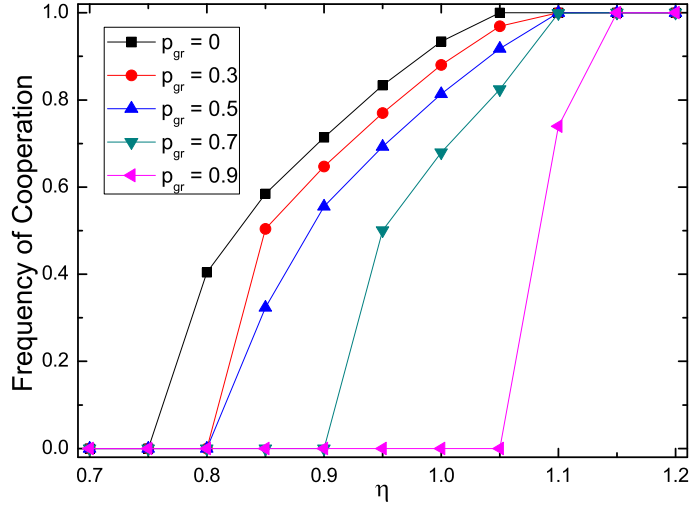}
\caption{The evolution of cooperation in spatial PGG with global redistribution of income as a function of $\eta$ with different values of $p_{gr}$.}
\label{GlobalPgrs}
\end{center}
\end{figure}

Similarly, we first show the evolution of the cooperative behaviours of the whole population as a function of $\eta$ with different $p_{gr}$ in \Fref{GlobalPgrs} to study the effect of global redistribution of income on the evolution of cooperation in spatial PGG.
It is well known that the goal of income redistribution is not to make all incomes equal by taking money away from some people and giving it to others in the whole population.
Instead, it is to avoid what proponents view as extreme or unreasonable inequality.
Thus, the case of $p_{gr} = 1.0$ is beyond the scope of our work.
Completely different from the previous results in \Fref{LocalPlrs}, the evolution of cooperation decreases with increasing $p_{gr}$ in the intermediate region of $\eta$ as shown in \Fref{GlobalPgrs}.
As we show above, the mechanism of local redistribution of income enhances the group heterogeneity on the square lattice and boosts the evolution of cooperation in spatial PGG.
While under the mechanism of global redistribution of income, we assume that the total payoff of each defector after $G$ PGGs is $\Pi_{gd}$, the total payoff of each cooperator is $\Pi_{gc}$.
Hence, the rest income of each defector after paying part of their income to the population according to the given probability $p_{gr}$ is $\Pi_{gd}'=\Pi_{gd}\cdot(1-p_{gr})$, and $\Pi_{gc}'=\Pi_{gc}\cdot(1-p_{gr})$ the rest income of each cooperator.
Then, the accumulated income is redistributed to the whole population irrespective of their strategies and contributions.
Accordingly, the final payoff of each defector is $\Pi_{gd}''=\Pi_{gd}'+S_{gG}/(L{\times}L)$, and $\Pi_{gc}''=\Pi_{gc}'+S_{gG}/(L{\times}L)$ the final payoff of each cooperator.
Consequently, the payoff difference between a cooperator and a defector is $\Pi_{gd}''-\Pi_{gc}''=\Pi_{gd}'-\Pi_{gc}'=(\Pi_{gd}-\Pi_{gc})\cdot(1-p_{gr})$.
Thus, the payoff difference between cooperators and defectors is not only determined by $1-p_{gr}$, but also the difference between $\Pi_{gd}$ and $\Pi_{gc}$.
With $\Pi_{gd}-\Pi_{gc}$ remaining unchanged, $1-p_{gr}$ helps to reduce the heterogeneity between cooperators and defectors of the whole population during the evolutionary process, which makes the inner incentive of players studying a random neighbour performing better decline.
Consequently, the evolutionary advantage of compact cooperative clusters cannot spread to the whole population.
Thus, the global redistribution of income actually inhibit the evolution of cooperation for the better.

\begin{figure}
\begin{center}
\includegraphics[scale=0.6]{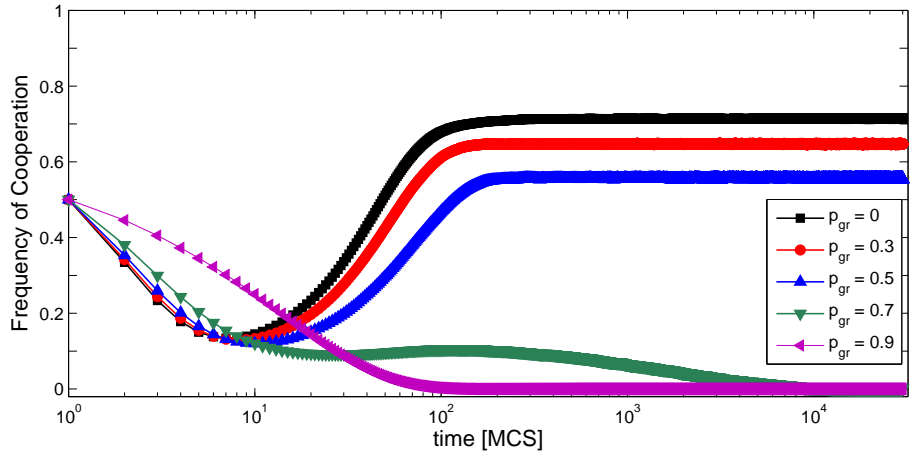}
\caption{
Time evolution as characterized by the frequency of cooperators with different $p_{gr}$.
Initially, cooperators and defectors are randomly distributed on the square lattice with equal probability.
The results are shown for $\eta=0.9$.
The evolutionary processes are averaged over 1000 different initial conditions.
}
\label{GlobalPgrTime}
\end{center}
\end{figure}

We also present the time evolution of cooperation in \Fref{GlobalPgrTime} to study the effect of the global redistribution of income intuitively.
We clearly demonstrate that the evolutionary dynamics of cooperation is fatigue with increasing $p_{gr}$.
Comparing with \Fref{LocalPlrTime} in the initial stage, the increment of $p_{gr}$ eases the fall of cooperation to some extent by narrowing the income gap between cooperators and defectors.
While in the long run, the mechanism of global redistribution of income reduces the heterogeneity between players resulting in the declination of cooperation level in the evolutionary steady state.
That is, the payoff diversity of players is gone.
Thereby, with high $p_{gr}$, players in the whole population have to pay more income to the whole population under this system except for the contribution to the common pool.
This makes people lose the motivation of pursuing a better life which should be the driving force of social progress.
In other words, there are more and more free-riders.
However, the mechanism of global redistribution of income do exist in real social-economic system, such as individual income tax.

\begin{figure}
\begin{center}
\includegraphics[scale=0.4]{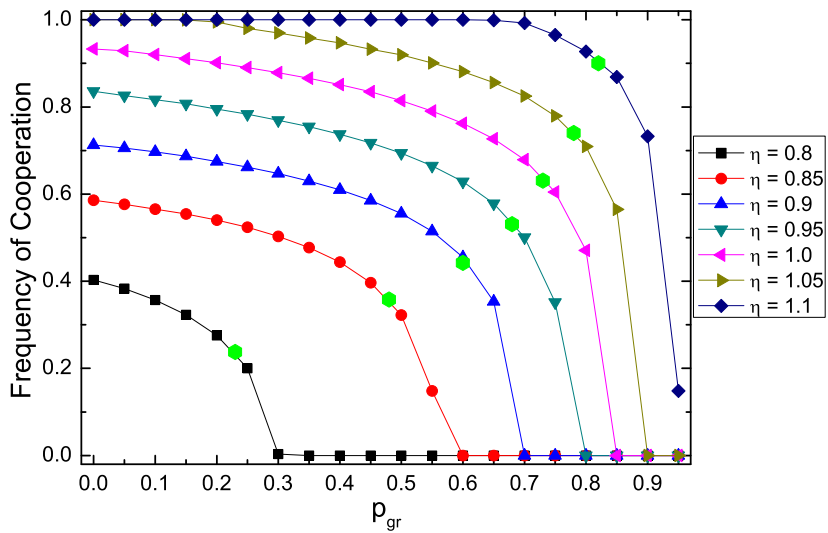}
\caption{
The evolution of cooperation in spatial PGG with global redistribution of income as a function of $p_{gr}$ with different values of $\eta$.
Green dots present the critical points of rapid decline for the frequency of cooperation on $\eta$s.
}
\label{GlobalEtas}
\end{center}
\end{figure}

In the following, we show \Fref{GlobalEtas} and deeply investigate the effect of global redistribution of income on the evolution of cooperation.
As is shown in \Fref{GlobalEtas}, the evolution of cooperation declines with increasing $p_{gr}$ on each fixed $\eta$ , which is consistent with \Fref{GlobalPgrTime}.
Remarkably, for each fixed $\eta$, the frequency of cooperation first declines slowly and then drops rapidly.
It is noted that the inflection point of a smooth curve is usually the maximum curvature point.
To point out the critical point of rapid decline for the frequency of cooperation, we calculate the maximum curvature of the frequency of cooperation for each fixed $\eta$, where the full cooperation and zero cooperation points are discarded.
It is found that there exists an intermediate maximum curvature point of $p_{gr}$ on each smooth curve with different $\eta$, which are marked as green dots in \Fref{GlobalEtas}.
For each fixed $\eta$, when $p_{gr}$ is smaller than the corresponding critical point, the cooperation level declines slowly.
While over the critical point of $p_{gr}$, the cooperation level plummets.
This phenomenon means that, on one hand, the mechanism of global redistribution of income reduces income inequality within the whole population, which is generally regarded to be a positive improvement to society.
But on the other hand, it may negatively affect the efficiency of social-economic system.
Thereby, the income expenditure proportion should be limited, which can be described as the social tolerance.
Beyond these limits, the social-economic system may be on the brink of collapse.
We argue that the stability of society is fundamental for the evolution of cooperation in social systems, which is essentially different from the models in biological systems\cite{Du2012,Du2014}.
In addition, we note that the critical points are closely related to $\eta$.

\begin{figure}
\begin{center}
\includegraphics[scale=0.4]{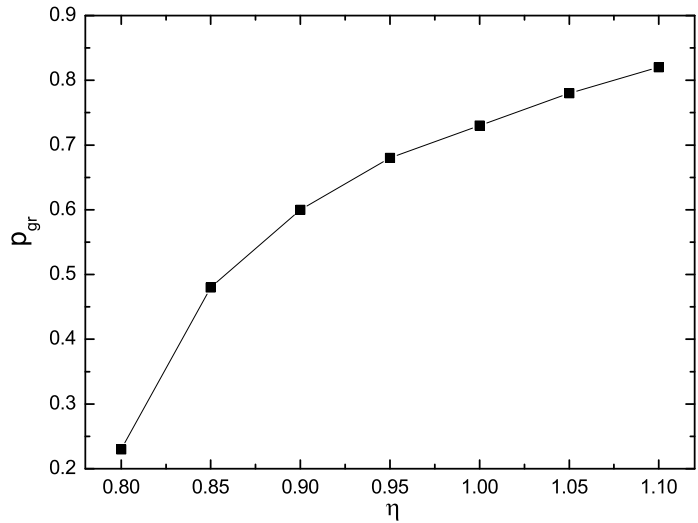}
\caption{
The corresponding relationship between $\eta$ and $p_{gr}$ in the evolution of cooperation with global redistribution of income.
}
\label{GlobalSocialTolerance}
\end{center}
\end{figure}

Social productivity affects the pre-tax income of players fundamentally\cite{Agranov2014}.
In this work, we assume that the renormalized enhancement factor $\eta$ denotes the social productivity.
In order to investigate the corresponding relationship between $\eta$ and $p_{gr}$ intuitively, we plot \Fref{GlobalSocialTolerance} to further study the effect of global redistribution of income on the evolution of cooperation.
Remarkably, there exists positive correlation between social tolerance and social productivity as shown in \Fref{GlobalSocialTolerance}.
With lower social productivity, where $\eta\rightarrow0.8$, lower income expenditure proportion is needed to rehabilitate the vulnerable cooperative behaviours.
While, when $\eta\rightarrow1.1$, the limited higher income expenditure proportion should be applied.
However, combining \Fref{GlobalEtas} where $\eta>0.95$, we point out that, once beyond the limit of critical point of $p_{gr}$, the social cooperation collapses more quickly with higher $\eta$.
On another level, when the social development is uneven, the progressive tax structure is necessary.
Low income people might prefer lower tax rates to reserve enough money for basic living.
People who make more money might pay higher taxes under this system to reduce social inequalities.
And, the result shown in \Fref{GlobalSocialTolerance} matches perfectly the property of tax progressivity\cite{Suits1977}.
This is the effectiveness and meaning of the redistribution of income.

\section{Conclusion and discussion}
In sum, we have studied the effects of redistribution of income on the evolution of cooperation in a spatial PGG.
Our research shows that local redistribution of income promotes cooperative behaviours, while global redistribution of income inhibits cooperation.
In the context of local income redistribution, income differences between cooperators and defectors in the same group are significantly influenced by parameter $p_{lr}$.
A group including more cooperators might be more successful in the evolution of cooperation\cite{Nowak2006}.
Although, group interactions may be less susceptible to the prevalence of cooperation owing to strong heterogeneity in the spatial public goods game\cite{Perc2011NJP}.
The mechanism of local income redistribution increases the possibilities of players learning from the members in cooperative groups, where the concept of group selection is continued.
While in the model of global income redistribution, payoff differences between players are eliminated with increasing $p_{gr}$.
The tendency of cooperation evolution is weakened by the reduction of payoff heterogeneity in the whole population.
Furthermore, there exist intermediate critical points of income expenditure proportion which indicate when social cooperation is maintained for different values of $\eta$.
It is also revealed that income expenditure proportion is positively associated with social productivity.
This highlights the subtle balance between social tolerance and social productivity.
That is, low social productivity restricts the efficient income expenditure proportion which retains sufficient income for the living of common people.
On the other hand, when social productivity is high, the limited higher income expenditure proportion can be applied to reduce social inequalities.
The right balance between them is subtle and difficult to pinpoint, which is associated with the specific social environment.
A qualitative analysis is provided in this work corresponding to the public goods dilemma.

Previous works on PGG have proposed a few mechanisms to boom cooperation.
Some literatures focused on the players' cognition of each other\cite{Vukov2012}.
Szab\'{o} and Szolnoki studied myopic players whose payoff interest is tuned from selfishness to other-regarding preference via fraternity adopting utility function\cite{Szabo2012}.
They found that the highest total income is achieved by the society whose members share their income fraternally.
Grund \etal distinguished between the evolution of individual preferences and behaviours by investigating self-regarding and other-regarding types of humans from an over-arching theoretical perspective to overcome the historical controversy in the behavioural sciences between largely incompatible views about human nature\cite{Grund2013}.
They demonstrated that a few ``idealists'' trigger off cooperation cascades, which can largely accelerate the spread of cooperation.
The above results are intriguing, but they suffer from a serious practical problem in social-economic system.
Actually, neither laying hopes on a few ``idealists'' nor over idealizing rock-solid brotherhood is practical.
In this paper, we continue along the concept of ``rational-economic man'', who makes decisions without considering the payoff or utility of others and the second-payment is mandatory.
Without endowing players with special characters, the effects of redistribution of income bring us lots of inspiration to the evolution of cooperation in real social-economic systems.

Some of former researches have studied how the presence of relatedness, which is incorporated into the game dynamics by redistributing a proportion of the payoffs to their immediate neighbours, affects the evolution of cooperation in the framework of the prisoner's dilemma game and the snowdrift game\cite{Wu2014}.
They found that the larger the proportion of the ultimate payoffs is contributed by the neighbouring individuals,
the more easily the cooperation can be established and persist in the population.
They argued that their work did not need an extra personal feature to characterize the other-regarding
preference of the individuals, and players make decisions without considering the payoff or utility of others.
Compared with these, our work provides a more profound understanding.
We have compared the characteristics on local redistribution of income and global redistribution of income, and studied their effects on the evolution of cooperation in spatial PGG.
We consider the real issues in a situation of ``networked minds'', which could significantly contribute to the converge of the behavioural sciences, and create a fundamentally new understanding of social-economic system.
With such simple and fundamental model, we could make more attempts to highlight the way to exploring the effects of social mechanisms on the evolution of cooperation in real society.

%Our results highlight the delicate effects of redistribution of income on the evolution of cooperation for the provisioning of public goods, and they reveal fascinating subtleties of the spatiotemporal dynamics in structured populations.

\ack{
We would like to thank L Yu, X J Chen and T Wu for their useful discussion and comments.
This work was supported by the National Natural Science Foundation of China (Grant No. 81500856) and Beijing Natural Science Foundation (Grant No. 7163215).
}

\end{document}